# Wireless Communications and Mobile Computing

# TLFW: A Three-layer Framework in Wireless Rechargeable Sensor Network with a Mobile Base Station

Anwen Wang[1#], Xianjia Meng[1#*], Lvju Wang[1], Xiang Ji[1], Hao Chen[2], Baoying Liu[1], Feng Chen[1], Yajuan Du[3], Guangcheng Yin[4]

[1] School of Information Science and Technology, Northwest University, Xi'an, China.
[2] China University of Labor Relations, Beijing, China.
[3] Beijing Zunguan Technology Company (National electronic computer quality supervision and Inspection Center).
[4] INSPUR SOFTWARE GROUP Co.,LTD .

[*] Correspondence should be addressed to Xianjia Meng; xianjiam@nwu.edu.cn

[#] Co-First author.

## Abstract

Wireless sensor networks as the base support for the Internet of things has been a large number of popularity and application. Such as intelligent agriculture, we have to use the sensor network to obtain the growth environmental data of crops, etc.. However, the difficulty of power supply of wireless nodes has seriously hindered the application and development of Internet of things. In order to solve this problem, people use low-power, sleep scheduling and other energy-saving methods on the nodes. Although these methods can prolong the working time of nodes, they will eventually become invalid because of the exhaustion of energy. The use of solar energy, wind energy, and wireless signals in the environment to obtain energy is another way to solve the energy problem of nodes. However, these methods are affected by weather, environment and other factors, and are unstable. Thus, the discontinuity work of the node is caused. In recent years, the development of wireless power transfer (WPT) has brought another solution to this problem. In this paper, a three-layer framework is proposed for mobile station data collection in rechargeable wireless sensor networks to keep the node running forever, named TLFW which includes the sensor layer, cluster head layer, and mobile station layer. And the framework can minimize the total energy consumption of the system. The simulation results show that the scheme can reduce the energy consumption of the entire system, compared with a Mobile Station in a Rechargeable Sensor Network(MSiRSN).

## I. Introduction

Internet of things (IoTs) are applied in everywhere now. Wireless sensor networks as the base support for the Internet of things has been a large number of popularity and application. Such as intelligent agriculture, we have to use the sensor network to obtain the growth environmental data of crops, etc.. However, in wireless sensor network, finite battery capacity is a major limitation of untethered nodes. Sensor nodes will operate for a finite duration, only as long as the battery lasts. the difficulty of power supply of wireless nodes has seriously hindered the application and development of Internet of things. In order to solve this problem, there are several solution techniques that have been proposed to maximize the lifetime of wireless sensor network, such as energy-aware routing protocols [1, 2], energy-efficient MAC protocols [3], redundant development of nodes [4], power management strategies [5, 6]. All the above techniques can maximize the lifetime of network. But the lifetime still remains bounded, and they will eventually become invalid because of the exhaustion of energy. The use of solar energy, wind energy, and wireless signals in the environment [7, 8] to obtain energy is another way to solve the energy problem of nodes. However, these methods are affected by weather, environment and other factors, and are unstable. Thus, the discontinuity work of the node is caused.

In recent years, the development of wireless power transfer (WPT)[9] has brought another solution to this problem. Wireless power transfer based on magnetic resonant coupling [10, 11] has been demonstrated to be a promising technology to address the problem in a wireless sensor network [12– 15]. In paper a Mobile Station in a Rechargeable Sensor Network(MSiRSN) [12], the authors showed how charging vehicle (WCV) can support wireless power transfer by bringing an energy source charge to proximity of sensor nodes and charging their batteries wirelessly, and carry the base station (MBS) to gather data. There is a home service station for the vehicle. The authors addressed the problem of co-locating the MBS on the WCV in a WSN by studying an optimization problem with a focus on the traveling path problem of the WCV, the data flowing routing depending on where the WCV is in the network, stopping points and charging schedule to minimize energy consumption of the entire system while ensuring none of the sensor nodes runs out of energy. In each charge period, WCV travels inside the network and charges every sensor node. In above papers, the traveling time is a little proportion of total time consisting of traveling time, vacation time and charging time. For example, the traveling time equals 1022s, the vacation time equals 10.26 hours and the charging time is 3.41 hours in the solution of simulation [13]. However, as the traveling time increases with the node density and the traveling time is a great part of the total time, traveling and charging every node in a period is improper.

To reduce the traveling time in a period, we propose a tiered system architecture consisting of the sensor layer, cluster head layer, and mobile station layer, as illustrated in Fig.1. The CBW travels all cluster heads at cluster head layer and selects the cluster in which CBW travels all sensor nodes at sensor layer in a sub-period. Several sub-periods form a period in which the sensor nodes in every cluster traveling once. Compared to traveling all nodes in [12, 13], the strategy reduces the proportion of the traveling time in total time, leading to reduction of the total energy consumption in the entire system, which includes power used by the CBW and the power consumed for wireless power transfer.

**Summary and contribution:**

• This paper designs a three-layer framework in rechargeable wireless sensor network with a mobile base station.

• A centralized clustering algorithm is proposed for sensors to organize them into m clusters and it will reduce energy consumption.

• This paper designs an optimization methods for joint charging schedule.

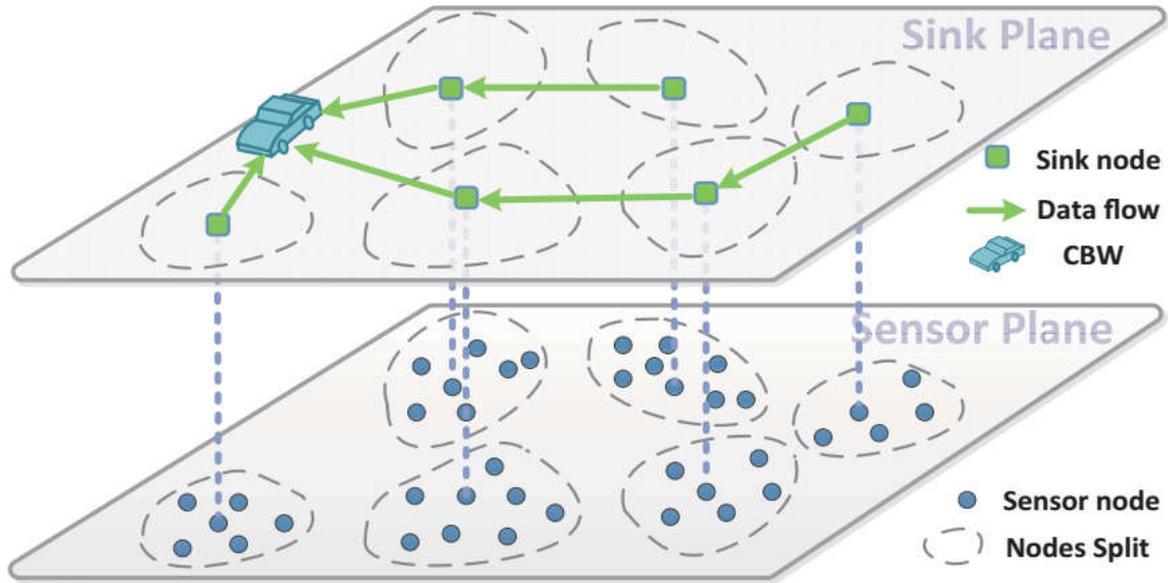

Fig. 1. Layered System Model

## II. OVERVIEW

In this paper we propose a tiered system architecture consisting of the sensor layer, cluster head layer and mobile station layer, to reduce the traveling time in a period. The CBW travels all cluster heads at cluster head layer and selects the cluster in which CBW travels all sensor nodes at sensor layer in a subperiod. Several sub-periods form a period in which the sensor nodes in every cluster traveling once. Compared to other methods, this strategy reduces the proportion of the traveling time in total time, leading to reduction of the total energy consumption in the entire system, which includes power used by the CBW and the power consumed for wireless power transfer.

At the sensor layer, a centralized clustering algorithm is proposed for sensors to organize them into m clusters and the sensor nodes transmit data to the cluster head via a single hop. In contrast to existing clustering methods which balance energy consumption, our scheme generates m cluster heads to minimize the total energy consumption. The single-hop data routing reduces energy consumption through that the sensor turns off the radio when there is no data generated by themselves to transmit.

At cluster head layer, cluster heads can cooperate with each other and the cluster head information is forwarded to CBW (a Car as Mobile bastion with Wireless power transfer) via multihop. The optimal flow routing is solved for a CBW moving trajectory to energy-saving.

At mobile station layer, we study an optimization problem that joints charging schedule for cluster heads and sensors, and flow routing for cluster heads.

## III. BACKGROUND

Wireless power transfer based on magnetic resonant coupling [10] has been demonstrated to be a promising technology to address the problem in a wireless sensor network in [12, 13]. In MSiRSN[12], the authors showed how charging vehicle (WCV) can support wireless power transfer by bringing an energy source charge to proximity of sensor nodes and charging their batteries wirelessly, and carry the base station (MBS) to gather data. There is a home service station for the vehicle. The authors addressed the problem of co-locating the MBS on the WCV in a WSN by studying an optimization problem with a focus on the traveling path problem [16] of the WCV, the data flowing routing depending on where the WCV is in the network, stopping points and charging schedule to minimize energy consumption of the entire system while ensuring none of the sensor nodes runs out of energy. In each charge period, WCV travels inside the network and charges every sensor node.

## IV. LAYERED NETWORK MODEL

In this section, we give an overview of entire framework. As depicted in Fig.1, it consists of three layers: sensor layer, cluster head layer and mobile station layer.

We consider a set of sensor nodes $N^*$ distributed over a two dimensional area. Each sensor node has a battery with a capacity of $E_{max}$ and the initial energy of battery is a random value. $E_{min}$ is denoted as the minimum level of energy at a battery for it to be operational. Each sensor node i generates sensing date with a rate $r_i$(in b/s),$i \in N^*$. Within the sensor network, there is a mobile CBW to charge sensor nodes and gather the entire network information.

In paper [12, 13], the authors proposed strategies to keep all nodes running forever using the wireless power transfer. In MSiRSN, the authors studied the problem of co-locating the MBS on the WCV in a WSN to minimize energy consumption of the entire system. The WCV follows a periodic schedule to travel inside the network for charging every sensor node.

However, as the traveling path increases with the number of sensor nodes, the time of traveling all sensor nodes is a large proportion of a period with large of sensor nodes in a wireless network. So traveling all sensor nodes in a period is an unwise strategy.

To address the issue, we introduce a three-layer model, consisting of the sensor layer, cluster head layer, and mobile station layer, as illustrated in Fig.1. The CBW travels all cluster heads N at cluster head layer and selects a cluster in which CBW travels all sensor nodes at sensor layer in a subperiod. Several sub-periods form a period in which the sensor nodes in every cluster traveling once. The schedule shows the cluster heads that with higher energy consumption have higher charging frequency than normal sensor nodes.

## V. CLUSTER SELECTION ALGORITHM

Since sensor nodes are energy-constrained, the network's lifetime is a major concern; especially for applications of WSNs in harsh environments. There are several solution techniques proposed, such as energy-aware routing protocols [1], energy-efficient MAC protocols [3], redundant development of nodes [4], power management strategies [5]. To support scalability of large WSN, nodes are often grouped into disjoint and mostly non-overlapping cluster. The most well-known hierarchical routing protocols are LEACH, HEED, TEEN, PEGASIS [17] etc. Other clustering algorithms in the literature varies in their objectives, such as load balancing [18], fault-tolerance [19], increased connectivity and reduced delay [20,32,33], minimal cluster count [21]. However, the above clustering algorithms are proposed to prolong the lifetime of network. Due to wireless power transfer technology, we can keep nodes running forever. So a cluster selection algorithm as following is proposed in order to minimize the entire energy instead of maximizing network longevity.

We propose a Algorithm to minimize the total communication energy consumption $C_s = \sum_{i=1}^{m} \sum_{t \in S_i} C_{ti} \cdot r_t (i \in \mathcal{N})$ in sensor layer. In order to clearly describe the algorithm, denote $\bar{N}$ as the number of elements in set N. The input of the algorithm are all node $N^*$ locations, communication consumption model parameters $\beta_1$, $\beta_2$ and $\omega$, and the number of cluster head m. The output of the algorithm are the m cluster in which there are several normal nodes and a cluster head to minimize the within-cluster sum of communication energy consumption. The algorithm have three step:

- First, randomly give initial cluster head set N, and $\bar{N} = m$.
- Secondly, we assign each node to a cluster that the node's communication energy consumption to the cluster head is minimum. The strategy yields the least within-cluster sum of communication energy consumption.
- Thirdly, we update the cluster head in a cluster through calculating the new mean to be the centroid of the sensor nodes in a cluster and setting the sensor nodes closest to the centroid as new cluster head in the cluster.
- Alternate second and third step until the centroids of all clusters do not change in range.

## VI. LAYER 1: NORMAL SENSOR NODES

### A. Static routing

We suppose every node's location is known in the network. By a clustering algorithm in section V, we can get m clusters and two type sensor nodes (normal nodes and cluster head nodes). To conserve the energy, we suppose the normal sensor nodes have not the collaboration capability, only send data to the cluster head via a single hop and do not forward packets coming from other sensor nodes.

The total data rate in cluster *i* (denoted as $R_i$) contains two parts, the first is data received from normal sensor nodes in cluster *i* and the second is the data generated by cluster head node *i*. Hence

$$R_i = \sum_{t \in S_i} r_t + r_i \quad (1)$$

Where $S_i$ is set of the type 0 nodes in cluster *i*, $r_i$ and $r_t$ is the data generation rate of node *i* and *t*, respectively. Given that cluster results, we have that $R_i$ is constant.

### B. Energy consumption in a cluster

Denote $C_{ti}$ as the energy consumption rate for transmitting one unit of data flow from normal sensor node *t* to cluster head *i*. Then $C_{ti}$ (in Joule/bit) can be modeled as [22]: $C_{ti} = \beta_1 + \beta_2 D_{ti}^\omega$ where $D_{ti}$ is the physical distance between node *t* and node *i*, $\beta_1$ and $\beta_2$ are constant terms, and $\omega$ is the path loss index and typically between $2 \leq \omega \leq 4$ [23].

$$D_{ti} = \sqrt{(x_t - x_i)^2 + (y_t - y_i)^2}$$

where $(x_t, y_t)$ and $(x_i, y_i)$ are the coordinates of cluster head $i$ and normal sensor node $t$. Given that all nodes are stationary and the cluster result is stationary, we have that $D_{ti}$, $C_{ti}$ are all constants.

Denote $\alpha$ (in Joule/bit) as the energy consumption rate for sensing one unit of data. The power consumption of the CPU is not taken into account.

For a normal sensor node $t$, the total energy consumption rate $c_t$, is

$$\alpha \cdot r_t + C_{ti} \cdot r_t = c_t, \quad (t \in S_i) \tag{2}$$

where $S_i$ is the set of normal nodes within the cluster $i$. Denote $C_i^*$ as the total energy consumption of all normal sensor nodes in a cluster $i$. Because the normal sensor nodes at a cluster only send data to the cluster head, there is no receiving energy consumption. Then we have

$$C_i^* = \sum_{t \in S_i} c_t \tag{3}$$

$$= \sum_{t \in S_i} (\alpha \cdot r_t + C_{ti} \cdot r_t), \quad (i \in \mathcal{N})$$

Given a cluster solution, the $C_i^*$ is constant. $C_s$ is denoted as the total energy consumption in the sensor layer, then we have:

$$\begin{aligned} C_s &= \sum_{i=1}^{m} C_i^* \\ &= \sum_{i=1}^{m} \sum_{t \in S_i} (\alpha \cdot r_t + C_{ti} \cdot r_t), \quad (i \in \mathcal{N}) \end{aligned} \tag{4}$$

where $m$ is the number of cluster.

### C. Charging model and charging behavior

In this section, we give a charging model and a charing behavior for normal sensor nodes in a cluster.

*1) Charging model:* Based on the charging technology [24], the vehicle with a wireless power transfer can charge neighboring nodes as long as they are within its charging range. We denote $U_{tB}(p)$ as the power reception rate at normal sensor node t when the vehicle position is $p$. Denote the efficiency of wireless charging by $\mu(D_{tB}(p))$ when the node is in charge range. $U_{max}$ is denoted as the maximum output power for a node and $D_\delta$ is denoted as the charging range of wireless power transfer. We assume power reception rate is too low to make magnetic resonant coupling work properly at the node battery, when the distance between the node and the mobile CBW.

wireless charging model is [13]:

$$U_{tB}(p) = \begin{cases} \mu(D_{tB}(p)) \cdot U_{max} & if \quad D_{tB}(p) \leq D_\delta \\ 0 & if \quad D_{tB}(p) > D_\delta \end{cases} \quad (5)$$

where $\mu(D_{tB}(p))$ is a decreasing function of $D_{tB}(p)$, and $0 \leq \mu(D_{tB}(p)) \leq 1$.

*2)    Cellular structure and energy charging behavior:* We consider all normal sensor nodes in a cluster. We employ the partition strategy in [13]. The two-dimensional plane of a cluster is partitioned into hexagonal cells with side length of *D*, as illustrated in Fig.2. We optional optimize the cell partition solution by the algorithm which the solution is illustrated in Fig.2.To charge normal sensor nodes in a cell, the mobile CBW only needs to visit the center of a cell. All normal sensor nodes within a hexagonal cell are within a distance of *D* from the cell center. Denote $Q_i$ as the set of all cell center and $q^*(q \in Q_i)$ as a cell. So the power reception rate of t in a cell with a center q is $U_{tB}(q) = \mu(D_{tB}(q)) \cdot U_{max}(q \in Q_i, t \in q^*)$, where $D_{tB}(q)$ is the distance between *t* and the center *q*. Note given the cell deployment and the $t^0$ position, the $D_{tB}(q)$ is constant. So we simple $D_{tB}(q)$ into $D_t$, $U_{tB}(q)$ into $U_t$.

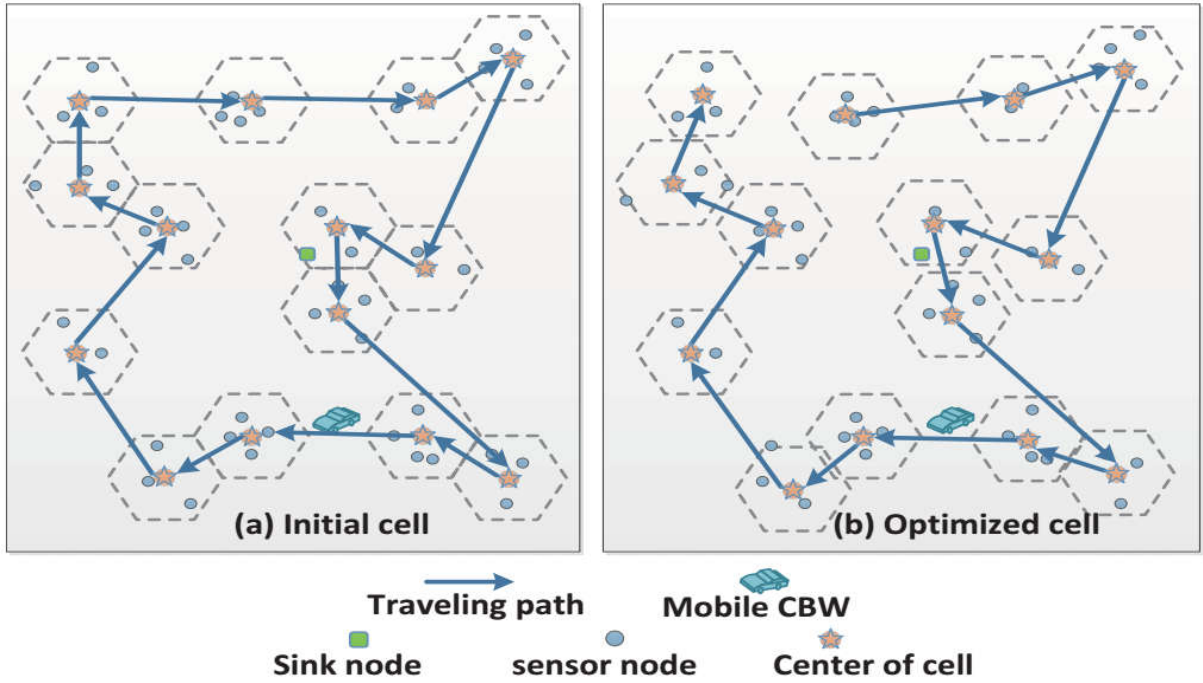

Fig. 2. Cell charging model at a cluster

We employ the so-called logical energy consumption rate $\gamma_t = \frac{c_t}{U_t}$ at a normal sensor node. Denote $\Gamma_i^*$ as the logical energy consumption rate at a cluster.

$$\begin{aligned}\gamma_t &= \frac{c_t}{U_t} \\ &= \frac{\alpha \cdot r_t + C_{ti} \cdot r_t)}{U_t}, \quad (t \in S_i)\end{aligned} \quad (6)$$

$$\begin{aligned}\Gamma_i^* &= \sum_{t \in S_i} \gamma_t \\ &= \sum_{t \in S_i} \frac{c_t}{U_t} \\ &= \sum_{t \in S_i} \frac{\alpha \cdot r_t + C_{ti} \cdot r_t)}{U_t}, \quad (i \in \mathcal{N})\end{aligned} \qquad (7)$$

*D. Traveling period in a cluster*

Denote P$_i$ as the traveling path and $\tau_i$ as the amount of time for each cycle. Then $\tau_i$ includes three components:

- The total traveling time along path P$_i$, D$_{Pi}$/$V$, where D$_{Pi}$ is the distance along path P$_i$ and $V$ is the traveling speed of the vehicle.
- The total sojourn time along path P$_i$, which is defined as the sum of all stopping time of the vehicle when it travels on P$_i$.
- The vacation time for the vehicle in a cluster $i$, $\tau_{vaci}$, which starts when the vehicle leaves the cluster $i$ and ends when the vehicle travels the path P$_i$ for charge all normal nodes in the cluster $i$.

Then we have:

$$\tau_i = \frac{\mathcal{D}_{\mathcal{P}_i}}{V} + \sum_{p \in \mathcal{P}_i}^{\omega_i(p) > 0} \omega_i(p) + \tau_{vac_i}, \quad (p \in \mathcal{P}_i) \qquad (8)$$

where $\tau_i(p)$ denotes the aggregate amount of time when the vehicle stays at point $p \in$ P$_i$ and $p_{vaci}$ denotes the vehicle leaves the cluster $i$ and is out of cluster $i$ charging period.

Note the mobile CBW only visits the cell center. To minimize the traveling time in a cluster, the mobile CBW must move along the shortest Hamiltonian cycle that connects the cluster head and the centers of cells in which there is at least one normal sensor node. The shortest Hamiltonian cycle can be obtained by solving the well known Traveling Salesman Problem (TSP) [25]. D$_{Pi}$ is denoted the solution of the TSP.

*E. Energy constrains for normal sensor node*

We offer two energy renewable conditions, and show that once they are met, the energy level at normal sensor node never falls below $E_{min}$ that means the normal sensor can run forever. First, we split energy consumption at normal sensor node $t$ in cluster $i$ into three part:

- energy consumed when the CBW do not select the cluster $i$ to charge battery: $\gamma_t \cdot \tau_{vaci}$.
- energy consumed when the CBW makes stops at all center of cell in which there is at least one normal sensor node: $\sum_{p \in \mathcal{P}_i}^{\omega_i(p) > 0, D_{tB}(p) > D_\delta} \gamma_t \cdot \omega(p)$.
- energy consumed when the CBW is moving along P$_i$ that is Hamiltonian cycle that connects the cluster head and the centers of cells in which there is at least one normal sensor node, $\gamma_t \cdot \frac{\mathcal{D}_{\mathcal{P}_i}}{V}$

$$E_{max} - [\gamma_t \cdot \tau_{vac_i} + \sum_{p \in \mathcal{P}_i}^{\omega_i(p)>0, D_{tB}(p)>D_\delta} \gamma_t \cdot \omega(p)$$
$$+ \gamma_t \cdot \frac{\mathcal{D}_{\mathcal{P}_i}}{V}] \geq E_{min}, \quad (i \in \mathcal{N}, t \in \mathcal{S}_i) \tag{9}$$

$$\gamma_t \cdot \tau_{vac_i} + \sum_{p \in \mathcal{P}_i}^{\omega_i(p)>0, D_{tB}(p)>D_\delta} \gamma_t \cdot \omega(p) + \gamma_t \cdot \frac{\mathcal{D}_{\mathcal{P}_i}}{V}$$
$$\leq \sum_{p \in \mathcal{P}_i}^{\omega_i(p)>0, D_{tB}(p)>D_\delta} U_{tB}(p) \cdot \omega(p), \quad (i \in \mathcal{N}, t \in \mathcal{S}_i) \tag{10}$$

We employ a cellular structure for normal sensor nodes (in section VI-C2). From (9) and (10), we obtain,

$$E_{max} - [\gamma_t \cdot \tau_{vac_i} + \sum_{q \in \mathcal{Q}_i} \gamma_t \cdot \omega(p)$$
$$+ \gamma_t \cdot \frac{\mathcal{D}_{\mathcal{P}_i}}{V}] \geq E_{min}, \quad (i \in \mathcal{N}, t \in \mathcal{S}_i) \tag{11}$$

$$\gamma_t \cdot \tau_{vac_i} + \sum_{q \in \mathcal{Q}_i} \gamma_t \cdot \omega(p) + \gamma_t \cdot \frac{\mathcal{D}_{\mathcal{P}_i}}{V} \tag{12}$$

$$\leq \sum_{p \in \mathrm{P}i}^{\omega i(p)>0, DtB(p)>D\delta} U_{tB}(p) \cdot \omega(p), \quad (i \in \mathrm{N}, t \in \mathrm{S}_i)$$

where $Q_i$ is the set of all cell center in cluster $i$.

## VII. LAYER 2: CLUSTER HEAD NODES

### A. Dynamic routing

With a clustering algorithm in section V to minimize the total energy of all sensor nodes, we can get some specific nodes and denote them as cluster heads. Different from normal nodes, the cluster head has collaboration capability.

### B. Dynamic flow balance

Due to the mobility of the vehicle, data flow routing is dynamic with routing topology changing over time. Denote $f_{ij}(p)$ and $f_{iB}(p)$ as flow rates from cluster head $i$ to cluster head $j$ and to the base station when the vehicle is at location $p \in P$, respectively. Then we have the following flow balance constraint at each cluster head $i$.

$$\sum_{k \in \mathrm{N}}^{k \neq i} fki(p) + Ri = \sum_{j \in \mathrm{N}}^{j \neq i} fij(p) + fiB(p) \ (i \in \mathrm{N}) \tag{13}$$

where N is the set of cluster heads gotten by cluster selection algorithm in section V, $R_i$ is determined by equation(1).

## C. Energy consumption

Like energy consumption model for normal sensor nodes, the communication energy consumption between two cluster nodes $i\ j$ can be modeled as:

$$C_{ij} = \beta_1 + \beta_2 D_{ij}^{\omega}$$

where

$D_{ij} = \sqrt{(x_i - x_j)^2 + (y_i - y_j)^2}$, $(x_i,y_i)$ and $(x_j,y_j)$ are the coordinates of cluster heads $i$ and $j$. $D_{iB}(p) = \sqrt{(x_i - x_B)^2 + (y_i - y_B)^2}$, $(x_i,y_i)$ and $(x_B,y_B)$ are the coordinates of type 1 node $i$ and vehicle $B$ at $p \in P$. Given that the cluster result and all cluster head are stationary, we have that $D_{ij}$, $C_{ij}$ are all constants. However $D_{iB}(p)$ and $C_{iB}(p)$ varied with vehicle position $p$. Denote $\gamma$ (in Joule/bit) as the energy consumption rate for receiving one unit of data. Then the total energy consumption rate for transmission, reception and sense at cluster head $i$ when the vehicle is at $p \in P$, denoted as $c_i(p)$, is

$$\alpha \cdot r_i + \rho \cdot \sum_{t \in S_i} r_t + \rho \cdot \sum_{k \in \mathcal{N}}^{k \neq i} f_{ki}(p) + X \sum_{k \in \mathcal{N}}^{j \neq i} C_{ij} \cdot f_{ij}(p)$$

$$+ C_{iB}(p) \cdot f_{iB}(p) = c_i(p) \quad (i \in P) \tag{14}$$

where $\alpha \cdot r_i$ is denoted as sensing consumption, $\rho \cdot \sum_{t \in S_i} r_t$ is denoted as consumption for receiving data from all normal sensor nodes at cluster $i$, $\rho \cdot \sum_{k \in \mathcal{N}}^{k \neq i} f_{ki}(p)$ is denoted as consumption for receiving data from other clusters head, $\sum_{k \in \mathcal{N}}^{j \neq i} C_{ij} \cdot f_{ij}(p)$ is denotes as consumption for transmitting data to other cluster, and $C_{iB}(p) \cdot f_{iB}(p)$ is denoted as consumption for transmitting data to the mobile CBW. Note the cluster head consumption $c_i(p)$ dynamically changes with the position $p$.

## D. Charging model

Like energy charging model for normal sensor nodes in section VI-C1, we use wireless power transfer [24] to charge the rechargeable battery of cluster heads. Different from normal sensor nodes charging schedule, the charging point for every cluster head is located in cluster head, taking account of the distance between any two cluster heads is longer than the charging range of wireless power transfer $D_\delta$, which means it is impossible to charge two cluster head simultaneously.

$$U_{iB}(p) = \begin{cases} U_{max} & \text{if} \quad p = i \\ 0 & \text{if} \quad p \neq i \end{cases} \tag{15}$$

where $p$ is the mobile CBW position, $p = i$ denotes the mobile CBW and cluster head $i$ are at the same position, and $p \neq i$ denotes the mobile CBW and cluster head $i$ are at two different positions. The equation (15). shows the mobile CBW just charges a cluster head while they are at the same position and do not charge battery when it is moving.

## E. Traveling period for cluster layer

Like section VI-D, the charge time for cluster heads is: $\tau = \frac{\mathcal{D}_\mathcal{P}}{V} + \sum_{p\in\mathcal{P}}^{\omega(p)>0} \omega(p) + \tau_{vac}, \quad (p \in \mathcal{P})$ where ω(p) denotes the aggregate amount of time the vehicle stays at point $p \in$ P and pvac denotes the location of the home service station. To minimize the traveling time of all cluster head, the mobile CBW must move along the shortest Hamiltonian cycle that connects the server station and all cluster head. Like traveling time at a cluster $D_P$, $D_P$ is denoted the solution of the TSP.

Based on the distance between two type 1 nodes is longer than charging range of wireless power transfer $D_\delta$, the points where the vehicle stops for charge cluster heads are in cluster heads position. Then the equation (8) can be written as follows:

$$\tau = \frac{\mathcal{D}_\mathcal{P}}{V} + \sum_{p\in\mathcal{P}}^{\omega(p)>0} \omega(p) + \tau_{vac}, \quad (p \in \mathcal{N}_i) \tag{16}$$

*F. Energy consumption in a sub-period*

We offer two energy renewable conditions, and show that once they are met, the energy level at clusters head never fall below $E_{min}$, which means the cluster head can run forever. First, we split energy consumption at normal sensor node *t* into three parts:

- energy consumed when the CBW makes a stop at service station): $c_i(p_{vac}) \cdot \tau_{vac}$
- energy consumed when the CBW makes stops at all cluster head: $\sum_{p\in\mathcal{N}, p\neq i} c_i(p) \cdot \omega(p)$,
- energy consumed when the CBW is moving along P that is Hamiltonian cycle that connects all cluster head and the service station, $\int_{s\in[0,D_P]}^{\omega(p(s))=0} \frac{1}{V} \cdot c_i(p(s))ds$

$$E_{max} - [c_i(p_{vac}) \cdot \tau_{vaci} + \sum_{p\in N, p\neq i} ci(p) \cdot \omega(p) \tag{17}$$

$$+ \int_{s\in[0,D_P]}^{\omega(p(s))=0} \frac{1}{V} \cdot c_i(p(s))ds] \geq E_{min}, \quad (i \in \mathcal{N})$$

$$c_i(p_{vac}) \cdot \tau_{vaci} + \sum_{p\in N, p\neq i} ci(p) \cdot \omega(p)$$

$$+ \int_{s\in[0,D_P]}^{\omega(p(s))=0} \frac{1}{V} \cdot c_i(p(s))ds$$

$$\leq \sum_{p\in\mathcal{P}}^{\omega(p)>0, D_{iB}(p)\leq D_\delta} U_{iB}(p) \cdot \omega(p), \quad (i \in \mathcal{N}) \tag{18}$$

## VIII. LAYER 3: CHARGING SCHEDULE AT CBW

We consider minimizing energy consumption of the entire system which includes normal sensor nodes and cluster heads. First, we minimize the total transmission energy consumption of all normal nodes in a cluster through a cluster selection strategy and give an optional charge strategy including an approximative optional path and charge time. Secondly, for cluster head layer, we formulate the problem including mobile CBW traveling path, dynamic flow routing and charge time, and solve the problem by CPLEX solver [ [26].

## A. Formulation for normal sensor nodes in a cluster

We develop a travel schedule for the mobile CBW, charging schedule among normal sensor nodes so that no normal node never runs out of energy. For the objective function, we consider minimizing energy consumption in sensor layer. We have following optimization problem (OPT-Normal).

$$max \frac{\tau_{vac_i}}{\tau_i}$$

s.t.  $Time constraints : (8);$

$Energy consumption model : (6),(7);$

$Energy renewable constraints : (11),(12);$

$\tau_i, \tau_{vac_i}, \omega_i(p) \geq 0 (p \in P_i);$

To minimize the traveling time in a cluster, the mobile CBW must move along the shortest Hamiltonian cycle that connects the cluster head and the centers of cells in which there is at least one normal sensor node. So $P_i$ is denoted the solution of this TSP.

## B. Formulation for cluster nodes

We develop a travel schedule for the mobile CBW, charging schedule and data flow routing among cluster heads so that no cluster head never runs out of energy. For the objective function, we consider minimizing energy consumption in cluster head layer. We have following optimization problem (OPT-Cluster)

$$max \frac{\tau_{vac}}{\tau}$$

s.t.  $Time constraints : (16);$

$Energy consumption model : (14);$

$Energy renewable constraints : (17),(18);$

$\tau, \tau_{vac}, \omega(p) \geq 0 (p \in P);$

$f_{ij}(p), f_{iB}(p), r_i(p) \geq 0 (i,j \in N, i \neq j, p \in P);$

To minimize the traveling time of all cluster head, the mobile CBW must move along the shortest Hamiltonian cycle that connects the server station and all cluster head. So $P$ is denoted the solution of the TSP.

## C. Joint solution

We find solutions to $D_{tps_i}$, $rate_i$ and $t_{max_i}$ for a cluster $i$, $\tau_{vac}$, $\tau$ and $\tau_{tsp}$ for cluster head by CPLEX [26]. Denote the $t_{max}$ as the $min(t_{max_i} \in N)$.

Denote *h* as the number of sub-period during which the mobile CBW charges all cluster heads once, in an entire period. Denote *T* as the total time of a period. Then we have $h \cdot \tau = T$. Denote $T_{vac}$ as the vacation time in the entire period. $T_{vac}$ equals the total time of a period minus the sum of the traveling time of cluster heads and normal sensor nodes, and the charging time of all nodes in the wireless network. $T_{vac}$ equals the total vacation time of cluster head traveling in *h* sub-period minus the sum of charging time and traveling path time of normal sensor nodes at all cluster. Then we have $h \cdot \tau - P{i \in N}(\Gamma*i \cdot T + Dtpsi/V) = Tvac$

To minimize the entire system consumption jointing sensor layer, cluster head layer and mobile bastion station, we study the following problem (OPT-Joint).

$max \frac{T_{vac}}{T}$

s.t.   $h \cdot \tau = T$;

$T \leq t_{max}$;

$h \cdot \tau_{vac} - \sum_{i \in \mathcal{N}}(\Gamma_i^* \cdot T + D_{tps_i}/V) = T_{vac}$;

$rate_i \cdot T + \frac{D_{tps_i}}{V} \leq \tau_{vac}, \quad (i \in \mathcal{N}$
$h > k$;                                        );

Constraint $T \leq t_{max}$ shows the entire period is no greater than the minimum value of maximum lifetime for a cluster in all clusters, which ensures every normal sensor node is not out of energy. $rate_i \cdot T + \frac{D_{tps_i}}{V} \leq \tau_{vac}, \quad (i \in \mathcal{N})$ shows the time that the mobile CBW is into a cluster and charges normal sensor nodes do not longer than the $\tau_{vac}$ that ensures in a sub-period, the mobile CBW can charge all normal sensor nodes. For the fractional objective function $\frac{T_{vac}}{T}$, we define $\eta_{vac} = \frac{T_{vac}}{T}$. Then we can reformulate above problem as following:

$max \eta_{vac}$

s.t.

$h \cdot \tau = T$;
$T \leq t_{max}$;
$h \cdot \eta_{vac} - \sum_{i \in \mathcal{N}}(\Gamma_i^* + \frac{D_{tps_i}}{V}/T) = \eta_{vac}$;
$rate_i \cdot T + \frac{D_{tps_i}}{V}/T \leq \eta_{vac}, \quad (i \in \mathcal{N})$;
$h > k$;

The equation $h \cdot \eta_{vac} - \sum_{i \in \mathcal{N}}(\Gamma_i^* + \frac{D_{tps_i}}{V}/T) = \eta_{vac}$ shows the $\eta_{vac}$ increases with the *T*. So we can maximize the $\eta_{vac}$ via maximizing *T* with two constraints $T \leq t_{max}$ and $\Gamma^*_i + \frac{D_{tps_i}}{V}/T \leq \tau_{vac}/T, \quad (i \in \mathcal{N})$, where $t_{max}$ and $\tau_{vac}/T$ can be calculated in section VIII-A and section VIII-A respectively. So the joint problem can be solved.

## IX. PERFORMANCE EVALUATIONS

In this section, we present some numerical results to demonstrate how our solution works to achieve wireless energy transfer and evaluate the performance of the system compared to MSiRSN.

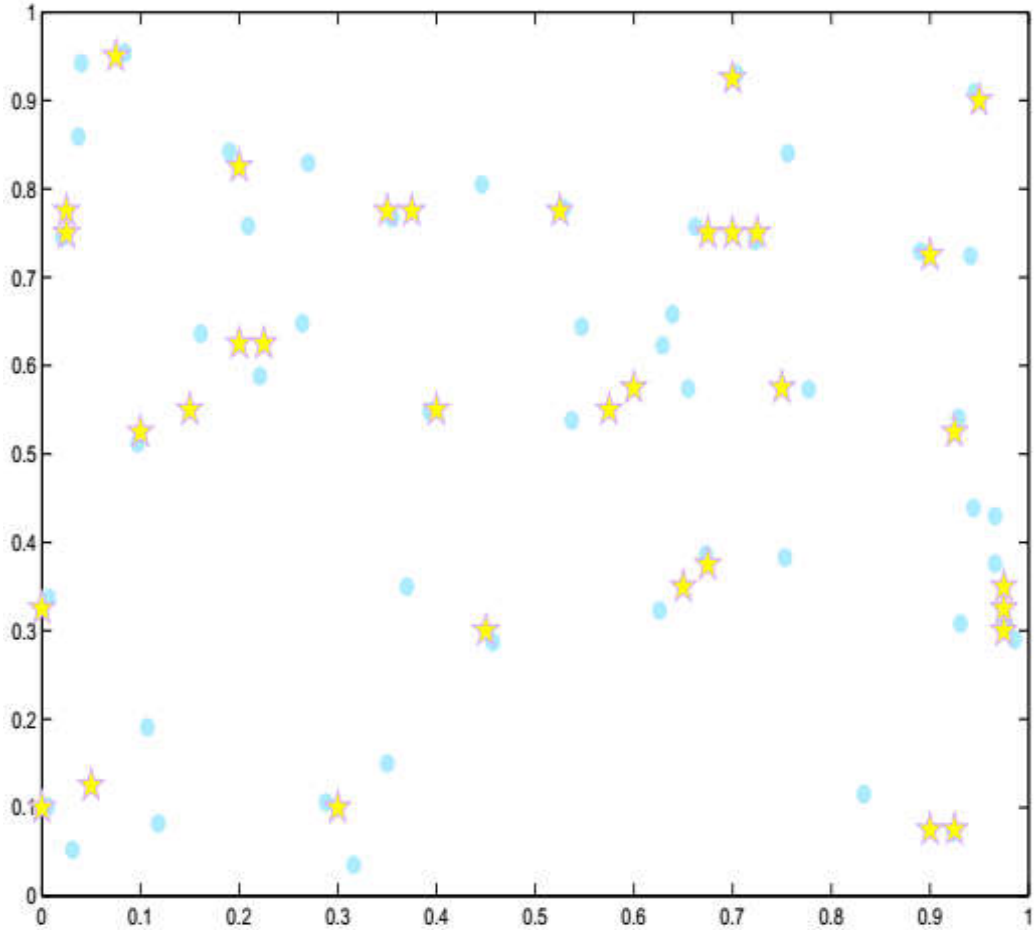

Fig. 3. The solution in MSiRSN. Yellow star represents stopping points, blue circle represents sensor node.

## *A. Simulation settings*

In this section, we evaluate the performance of the system and compared it with the strategy in MSiRSN. The network parameters are set like in MSiRSN. We assume sensor nodes are deployed over a 1×1 square area. The service station is at (0.5,0.5). The traveling speed of the mobile CBW is $V = 0.1$. The data rate $r_t$, $t \in N^*$, from each node is randomly generated within [0.1,1]. Power consumption coefficients are $\beta_1 = 1, \beta_2 = 1, \rho = 1, \alpha = 0$. The path loss index is $\omega = 4$. Suppose that a sensor nodes uses a rechargeable battery with $E_{max} = 10,000$, and $E_{min} = 500$. For the charging efficiency function $\mu(D_{tB}(p)) = -40D_{tB}(p)^2 - 4D_{tB}(p)^2 + 1$. Letting $U_{max} = 50$, $D_\delta = 0.1$ for a maximum distance of effective charing. We consider a 50-node network. The normalized location of each node and its data rate is given at Table 1 in MSiRSN.

Table 1: Location and data rate Ri for each node in a 50-node network

| Node Index | Location | $R_i$ | Node Index | Location | $R_i$ |
|---|---|---|---|---|---|
| 1 | (0.547,0.644) | 0.1 | 26 | (0.833,0.115) | 0.2 |
| 2 | (0.662,0.757) | 0.7 | 27 | (0.639,0.658) | 0.1 |
| 3 | (0.037,0.859) | 0.4 | 28 | (0.704,0.930) | 0.6 |
| 4 | (0.723,0.741) | 1.0 | 29 | (0.977,0.306) | 0.8 |
| 5 | (0.529,0.778) | 0.9 | 30 | (0.673,0.386) | 0.5 |
| 6 | (0.316,0.035) | 0.4 | 31 | (0.021,0.745) | 0.7 |
| 7 | (0.190,0.842) | 0.8 | 32 | (0.924,0.072) | 0.6 |
| 8 | (0.288,0.106) | 0.8 | 33 | (0.270,0.829) | 0.1 |
| 9 | (0.040,0.942) | 0.2 | 34 | (0.777,0.573) | 0.8 |
| 10 | (0.264,0.648) | 0.4 | 35 | (0.097,0.512) | 0.9 |
| 11 | (0.446,0.805) | 0.5 | 36 | (0.986,0.290) | 0.2 |
| 12 | (0.890,0.729) | 0.5 | 37 | (0.161,0.636) | 0.7 |
| 13 | (0.370,0.350) | 0.1 | 38 | (0.355,0.767) | 0.9 |
| 14 | (0.006,0.101) | 0.7 | 39 | (0.655,0.574) | 0.5 |
| 15 | (0.393,0.548) | 0.1 | 40 | (0.031,0.052) | 0.4 |
| 16 | (0.629,0.623) | 0.1 | 41 | (0.350,0.150) | 0.3 |
| 17 | (0.084,0.954) | 0.5 | 42 | (0.941,0.724) | 0.1 |
| 18 | (0.756,0.840) | 0.2 | 43 | (0.966,0.430) | 0.2 |
| 19 | (0.966,0.376) | 0.7 | 44 | (0.107,0.191) | 0.3 |
| 20 | (0.931,0.308) | 0.6 | 45 | (0.007,0.337) | 0.3 |
| 21 | (0.944,0.439) | 0.1 | 46 | (0.457,0.287) | 0.4 |
| 22 | (0.626,0.323) | 0.4 | 47 | (0.753,0.383) | 0.1 |
| 23 | (0.537,0.538) | 0.2 | 48 | (0.945,0.909) | 0.1 |
| 24 | (0.118,0.082) | 0.3 | 49 | (0.209,0.758) | 0.3 |
| 25 | (0.929,0.541) | 0.2 | 50 | (0.221,0.588) | 0.8 |

## B. Solution with strategy in MSiRSN

The simulation results in MSiRSN are given as following. The traveling path in a period $D_{POPT-lb} = 4.89$ and the traveling time $D_{POPT-lb}/V = 48.9$. The cycle time $\tau = 9414$, the vacation time $\tau_{vac} = 6410$, and the objective value = 68%. The traveling path is shown in Fig.3.

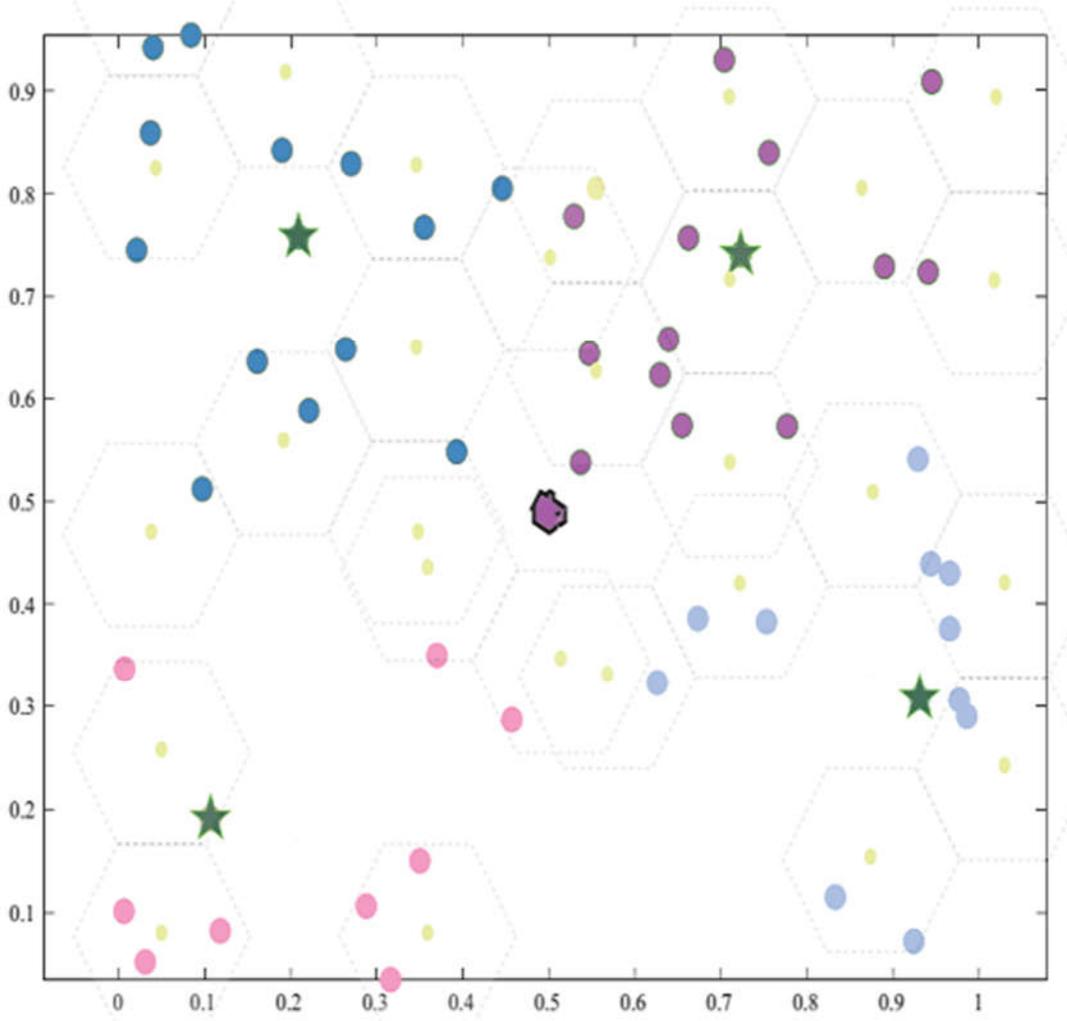

Fig. 4. The layer framework of a network in section IX. Green star represents cluster head, yellow ellipse represents cell center that is stopping point for charging normal nodes in the cell, other four colour circles represent four types normal nodes.

## C. Solution with our strategy

With our strategy, we can get a layer framework of the network, shown in Fig.4. We solve optimization problems (OPT-Normal and OPT-Cluster) by CPLEX [26] and get following solutions. Denote $\tau_c$ as the total charging time for cluster heads in a sub-period, $\tau_{ci}$ as the total charging time for normal sensor nodes at cluster $i$.

- $D_{tps}$ = 2.5425, $\tau_{vac}$ = 209.1, $\tau_c$ = 52.3, $\tau$ = 286.8
- $D_{tps1}$ = 1.3196, $\tau_{vac1}$ = 10674, $\tau_{c1}$ = 30, $\tau_1$ = 10717.
- $D_{tps2}$ = 1.239, $\tau_{vac2}$ = 8703, $\tau_{c2}$ = 56, $\tau_2$ = 8771.3.
- $D_{tps3}$ = 1.5124, $\tau_{vac3}$ = 8416.7, $\tau_{c2}$ = 89, $\tau_3$ = 8520.
- $D_{tps4}$ = 1.732, $\tau_{vac4}$ = 5946.57, $\tau_{c2}$ = 171, $\tau_4$ = 6134.

We set a period $T = \tau_4 = 6134$. We can get charging time for normal nodes $P_{\tau ci}$ =17+39+64+171=291, charing time for cluster heads $\tau_c \cdot T/\tau$ =1115, traveling time for normal nodes $^P D_{tpsi}/V$=84, traveling time for cluster heads $\frac{D_{tps}}{V} \cdot T/\tau = 543$ in this period. Then we can get $\frac{T_{vac}}{T} = \frac{6134-(291+1115+84+543)}{6134} = 0.71$ Our objective solution 71% is greater than 68% in MSiRSN.

## X. RELATED WORK

The lifetime of wireless sensor networks is often limited by energy supplies. The problem of node energy supply is also a key problem in the application development of wireless sensor networks. To solve this problem, researchers have explored a wide variety of solutions.

One type of the methods is to save energy by optimizing the hardware and software [27] of the nodes. There are several solution techniques that have been proposed to allow nodes to work as long as possible in a limited amount of energy. Such as energy-aware routing protocols [1], energy-efficient MAC protocols [3], redundant development of nodes [4], power management strategies [5].But no matter how energy-efficient, the battery will eventually be used up. Then the network is invalid.

Another type of the methods is to automatically obtain energy by nodes from the natural environment. Such as wind, solar energy [28], etc.. The energy harvesting techniques referring to har-nessing energy from the environment and converting energy to electrical energy make that a node can be powered perpetually possible, such as [7, 8]. Due to uncontrollability and unpredictability of the energy source that refers to the ambient source of energy to be harvested, the techniques can not ensure that nodes run in every moment.

The third type of the methods is to obtain energy using ubiquitous radio signals [29]. However, this technology is still in its initial stage of research, and can obtain very little energy. This is mainly caused by the far distance and the limited transmitting power of the electromagnetic wave. Recently, wireless power transfer based on magnetic resonant coupling [10] has been demonstrated to be a promising technology to address the problem in a wireless sensor network [12–14, 30, 31-40]. In MSiRSN, the authors showed how charging vehicle (WCV) can support wireless power transfer by bringing an energy source charge to proximity of sensor nodes and charging their batteries wirelessly. But the overall energy consumption is higher.

In this paper, we propose a tiered system architecture consisting of the sensor layer, cluster head layer, and mobile station layer to reduce the traveling time in a period. The CBW travels all cluster heads at cluster head layer and selects the cluster in which CBW travels all sensor nodes at sensor layer in a sub-period. Several sub-periods form a period in which the sensor nodes in every cluster traveling once. Compared to traveling all nodes in above methods[12, 13], the strategy reduces the proportion of the traveling time in total time, leading to reduction of the total energy consumption in the entire system, which includes power used by the CBW and the power consumed for wireless power transfer.

## XI. CONCLUSION

Wireless sensor networks is the main part of IoTs. With the high developing time of IoTs, the difficulty of power supply of wireless nodes has seriously hindered the application and development of IoTs. In this paper, we proposed a threelayer framework consisting of the sensor layer, cluster head layer, and mobile station layer in a rechargeable wireless sensor network. We studied the problem of charge schedule and traveling path of a mobile CBW and a cluster selection algorithm in order to minimize the energy consumption of entire system. The simulation result shows that the scheme can get a smaller the energy consumption of the entire system, compared with MSiRSN.

## Data Availability

The data used in this paper can be obtained directly in the sentences and tables of the paper or generated by combining them with the algorithm. The core steps and algorithms of data processing method are introduced in the paper in detail, too.

## ACKNOWLEDGMENT

This work was Supported in part by Scientific Research Program Funded by Shaanxi Provincial Education Department (Program No. 17JK0775 and 18JK0773), and the Science Foundation of Northwest University (no.15NW32 and no.15NW31), and the Natural Science Basic Research Plan in Shaanxi Province of China (no. 2017JM6056 and 2017GY-191). This work is also supported in part by the Key Research and Development Program of Shaanxi Province (No. 2018ZDXM-GY-036).